\documentstyle[12pt]{article}
\begin{document}

\begin{flushright}
OUTP-96-17P\\
hep-ph/9610298\\
March 1996
\end{flushright}
\vskip 1.5 cm
\centerline{\large\bf Simple GUTs\footnote{Presented 
as plenary talk at {\it QUARKS96}, Yaroslavl, Russia,
May 5-11, 1996. To be published
in the proceedings.}}
\vskip 2 cm
\begin{center}
{\bf Giovanni AMELINO-CAMELIA}\\
\end{center}
\begin{center}
{\it Theoretical Physics, University of Oxford,
1 Keble Rd., Oxford OX1 3NP, UK}
\end{center}
\vskip 1 cm
\centerline{\bf ABSTRACT }
\medskip
Together with some positive
features,
such as the constrained hypercharge assignments,
grandunification models
typically have some unwelcome aspects,
mostly associated with the rich and peculiarly selected
matter content.
I discuss some ideas which might help to interpret
(part of) this apparent complexity
arising in grandunification in terms of simple structures.
I also point out that the conventional 
criteria used to establish fine tuning
in grandunified models
should be modified if
these models 
are viewed as effective
low-energy descriptions of a more fundamental theory.


\section{Introduction}

$~~~~~~$In spite of its remarkable phenomenological success,
the Standard Model ($SM$) is not believed
to provide a fundamental theory of particle physics.
This expectation originates for many of us
not only from the fact that $SM$ does not incorporate
gravity, and should therefore become
inadequate at some very high scale\footnote{At this high scale,
possibly the Planck mass $M_P \! \sim \! 10^{19} GeV$, 
one must after all even consider
the possibility that the usual tools of particle physics
might loose meaning.},
but also because of its rather complex structure.
It is probably worth emphasizing, at least for the benefit
of the students who attended Quarks96 and {\it might} read the proceedings,
that we have nothing (from the conceptual
viewpoint) to assure us that nature should be describable
in terms of simple laws. Still, most of us do expect this simplicity,
probably extrapolating from the history of physics,
which has proceeded through a series of steps
of deeper understanding and simplification (such as the description
of the baryon spectrum in terms of the quark model).
From the point of view of this expected simplicity,
the $SM$ is quite unsatisfactory, since it leaves {\it unanswered}
several questions;
in particular,

\begin{itemize}
\item[(Qa)] Particle physics is described
by a gauge theory with the peculiar gauge
group $G_{SM} \! \equiv \! SU(3)_c \otimes SU(2)_L \otimes U(1)_Y$.
\item[(Qb)] The corresponding three coupling
constants, $\alpha_s$, $\alpha_2$, and $\alpha_Y$,
are free parameters of the model.
\item[(Qc)] A peculiar bunch of IRREPs (irreducible representations)
of $SU(3)_c \otimes SU(2)_L$ hosts the quarks and leptons.
\item[(Qd)] The hypercharge assignments to the quark and lepton IRREPs
are arbitrary.
\item[(Qe)] The Yukawa couplings are arbitrary.
\item[(Qf)] The entries of the Cabibbo-Kobayashi-Maskawa
matrix are arbitrary.
\item[(Qg)] Each of the quarks and leptons of the model
is present in triplicate copy (family structure).
\item[(Qh)] A peculiar (bunch of) IRREP(s) hosts the Higgs particles.
\item[(Qi)] The parameters of the Higgs potential,
which determine the Higgs mass(es) and all the aspects of
symmetry breaking, are arbitrary.
\item[(Qj)] The anomaly cancellation is a (apparently accidental) result
of the structure 
of the (arbitrarily selected) matter content of the model.
\end{itemize}

GUTs [models with a (grand)unified description of particle interactions]
have been investigated primarily because,
as illustrated in the discussion of $SO(10)$ GUTs given
in the next section, they address/simplify (Qa)-(Qf) and,
in some cases, (Qj),
and are therefore good candidates for the description of
particle physics if the trend of incremental simplifications
of this description
is to continue.
However, it should be noted that not only GUTs bring no improvement
in relation to (Qg), but they actually increase the complexity
associated to (Qh) and (Qi).
Therefore, from the ``aesthetic'' viewpoint,
GUTs have merits and faults (with the merits outnumbering,
but not necessarily outweighing, the faults).

Phenomenological encouragement for the GUT idea comes from the observed
low-energy values of $\alpha_s$, $\alpha_2$, and $\alpha_Y$,
which appear to be arranged just as needed
for unification.
Indeed, (although the simplest GUT candidate, minimal $SU(5)$,
does not pass this test\cite{amal})
there are several examples of GUTs which reproduce
these data on the coupling constants while
being consistent with 
the present\cite{padb} experimental lower limit 
on proton decay, $\tau_{p\rightarrow e^+\pi^0} \geq 9\cdot10^{32}~years$.
One important feature of phenomenologically consistent GUTs
is that they must involve at least one extra scale,
besides the unification scale $M_X$,
at which the RGEs (renormalization group equations)
of the $SM$ couplings are modified.
In fact,
the recent precise determination of the gauge coupling constants at 
the scale $M_Z$ has allowed to show that, if only $SM$ particles 
contribute to the RGEs between $M_Z$ and $M_X$, the
three running coupling constants of $SM$
meet at three different points\cite{amal} 
and only the meeting 
point of $\alpha_2(\mu)$ and $\alpha_s(\mu)$
corresponds to a value of the scale 
$\mu$ sufficiently high to comply with
the experimental lower limit on proton decay.
Equivalently, one can describe this situation by stating that if 
only $SM$ particles 
contribute to the RGEs between $M_Z$ and $M_X$, a
one-step unification would require low-energy 
values of the couplings that are inconsistent with experimental data.

Perhaps the simplest GUTs
meeting the minimum requirement of
agreement with the data on the coupling constants and
with
the experimental limit on proton decay
are some $SO(10)$ models, which are reviewed in the next section.
They naturally (see next section)
predict a two-scale breaking to $G_{SM}$;
in fact, a typical $SO(10)$ breaking chain is given by
{\small
\begin{equation}
SO(10)\buildrel M_X\over \longrightarrow G'\buildrel M_R\over
\longrightarrow
G_{SM} \buildrel M_Z\over\longrightarrow
SU(3)_c \otimes U(1)_{e.m.}
\end{equation}
}
Importantly, in $SO(10)$ the hypercharge $Y$ is the combination 
of two generators belonging to the Cartan,
\begin{equation}
Y=T_{3R}+\frac{B-L}{2}
~,
\end{equation}
where 
$B-L$ and $T_{3R}$ 
belong respectively 
to the 
$SU(4)_{PS}$
(the $SU(4)$ containing $SU(3)_{c}$ and $U(1)_{B-L}$,
which was first considered
by Pati and Salam) and the $SU(2)_R$ subgroups of $SO(10)$.
This leads to a high unification point $M_X$
if the intermediate symmetry 
group $G'$ contains $SU(2)_R$ and/or $SU(4)_{PS}$, since
then, between $M_Z$ and $M_R$,
the Abelian evolution of $\alpha_Y$ (predicted by $SM$)
is replaced by 
the non-Abelian one of either component of $Y$. $M_X$ is connected
with the masses of the lepto-quarks that can
mediate proton decay, and 
this $SO(10)$ mechanism
for a higher unification point 
turns out to be useful in allowing to meet the condition
\begin{equation}
M_X \geq 3.2\cdot10^{15}~GeV
~, \label{eq:mxinf}
\end{equation}
\noindent 
which is necessary\cite{oliv} for agreement
with the present experimental
limit on proton decay.

Although (relatively) simple GUTs, such as $SO(10)$, {\it can work}, 
they are affected by the hierarchy
problem,
and this causes many to prefer SUSY (supersymmetric) GUTs.
Also model building in the SUSY GUT case
has little difficulties reaching
the minimum requirement of
agreement with
the data on the coupling constants and
the experimental limit 
on proton decay;
SUSY $SU(5)$ GUTs already meet
this minimum requirement.
These models typically predict one-step breaking
of $SU(5)$ to $G_{SM}$ at the scale $M_X$,
but the RGEs are already modified at a much lower scale, $M_{susy}$,
where SUSY breaking occurs

{\small
\begin{equation}
SUSYSU(5) \buildrel M_X\over\longrightarrow
SUSYG_{SM} \buildrel M_{susy} \over \longrightarrow
G_{SM} \buildrel M_Z\over\longrightarrow
SU(3)_c \otimes U(1)_{e.m.}
~ \label{eq:ssb}
\end{equation}
}

In this paper (but not necessarily elsewhere) I take the position
that for the GUTs (whether they are SUSY or not)
the aesthetic advantages and the consistency with
the observed low-energy values of the gauge coupling constants
outweigh the {\it damage done} in regard to (Qh) and (Qi).
This motivates me to look for possible ways to associate
hidden simplicities to
the apparently complicated GUT structures
affecting (Qh) and (Qi) (and (Qg)); the reader is warned 
of the fact that 
the resulting discussion is accordingly quite speculative.

\section{$SO(10)$ Models as GUT Prototypes}

$~~~~~~$In this section I illustrate some of the aspects of GUTs
that I mentioned
in the Introduction, using as an example the GUTs based on the $SO(10)$
group\cite{frmi}.
In these models the electromagnetic, weak, and strong interactions
are unified in an $SO(10)$ gauge interaction characterized
by one gauge coupling.
$SO(10)$ GUTs also reduce the amount of arbitrariness which 
characterizes the $SM$
in regard to the content of leptons and quarks;
in fact, they accommodate in one IRREP,
the 16-dimensional spinorial representation, the fifteen known 
left-handed fermions of a generation plus a new particle, whose 
quantum numbers are the same as those of the not-yet-discovered $\nu_L^c$.
Within the 16 (here and in the following I refer to the IRREPs of
a group by their dimension only), quarks and leptons are also
automatically assigned hypercharges in agreement with the observed
quantization; these hypercharge assignments are instead a
free input of $SM$.
Moreover, the presence of $\nu_L^c$ can lead to a mass matrix, for one 
generation of neutrinos, of the form
\begin{eqnarray}
  \begin{array}{lll}
    \begin{array}{lr}
      \left(\overline{\nu_R^c} \right. & \left.\overline{\nu_R}\right) \\
                                      &
    \end{array}\!\!\!\!\!
    &
    \!\!\!\!\left(
    \begin{array}{lr}
      0 & \frac{m_D}{2} \\
      \frac{m_D}{2} & M
    \end{array} 
    \right)\!\!\!
    &
    \!\!\!\left(
    \begin{array}{c}
    \nu_L \\
    \nu_L^c
    \end{array}
    \right)
  \end{array}\!\!\! +~h.c.,
\label{eq:masmat}
\end{eqnarray}
where $m_D$ is a Dirac mass, which can be conjectured to be
of the order of the other masses of the fermions in the given generation,
and $M$ is 
a Majorana mass for the right-handed
neutrinos. If $m_D \! \ll \! M$, 
this mass matrix
has the two eigenvalues
\begin{equation}
m_{\nu_1}\sim \frac{m_D^2}{M},~~m_{\nu_2}\sim M,
\end{equation}
leading to the relation $m_{\nu_1}\ll m_D$ as observed experimentally.

Interestingly, the choice of $SO(10)$ as the grandunification 
group also leads to
the absence of triangle anomalies, due to the fact that in $SO(10)$ 
it is not possible to construct a cubic invariant with the adjoint 
representation, which describes the gauge bosons.
[In the $SM$, and in some GUTs,
there is an accidental ({\it ad hoc}) cancellation among
different (nonvanishing) contributions to the anomaly.]

Up to this point I have only discussed positive features of $SO(10)$
GUTs (and ignored the annoying
family triplication, which is left unmodified by
the move from $SM$ to $SO(10)$, and in $SO(10)$
is understood to imply
the need for three copies of the 16).
This has been the case because I have not yet
discussed the Higgs sector.
I shall do this next. It will
put in evidence several puzzles,
while allowing me to point out that indeed, as needed for the
phenomenological considerations
mentioned in the Introduction,
in $SO(10)$ GUTs it is quite natural to have a two-step
breaking to $SM$.

In discussing the Higgs sector it is useful to
classify the components of the 
smallest IRREPs of the group that are invariant under $G_{SM}$.
The IRREPs 10,120,320 contain no $G_{SM}$ singlet.
The 16 and the 126 contain one $G_{SM}$ singlet each,
and in both cases the little group of the singlet is $SU(5)$.
The 45 contains two independent $G_{SM}$ singlets
({\it i.e.}, it contains a two dimensional space of $G_{SM}$ singlets), 
one with little group
$SU(3)_c \otimes SU(2)_L \otimes SU(2)_R \otimes U(1)_{B-L} \times D$
(D stands for the 
left-right discrete symmetry\cite{kuzm} 
which interchanges $SU(2)_L$ and $SU(2)_R$)
and the other with 
little group $SU(4)_{PS} \otimes SU(2)_L\otimes U(1)_{T_{3R}}$.
The 54 contains one $G_{SM}$ singlets, with little group
$SU(4)_{PS} \otimes SU(2)_L \otimes SU(2)_R \times D$.
The 144 contains one $G_{SM}$ singlet, with little group $G_{SM}$.
The 210 contains three independent $G_{SM}$ singlets,
one with little group
$SU(3)_c \otimes SU(2)_L \otimes SU(2)_R \otimes U(1)_{B-L} \times D$,
one with little group 
$SU(4)_{PS} \otimes SU(2)_L \otimes SU(2)_R$,
and one
with little group 
$SU(3)_c \otimes SU(2)_L \otimes U(1)_{T_{3R}} \otimes U(1)_{B-L}$.

From these properties of the
smallest IRREPs of $SO(10)$
one can easily see that the typical pattern
of SSB of $SO(10)$ to $G_{SM}$
has two steps; indeed, 
with the exception
of the singlet in the 144, 
the little group of
all the
above mentioned $G_{SM}$ singlets
is larger than $G_{SM}$.

Actually, either for phenomenological or for technical reasons,
some of the 
above mentioned $G_{SM}$ singlets,
cannot be used for the first SSB step.
The use of the $G_{SM}$ singlets in the
16, 126 and 144 representations would lead to 
the result that, like in $SU(5)$ GUTs,
the unification 
of the $G_{SM}$ coupling constants 
is inconsistent
with the low-energy 
data on these couplings.
(In the cases of the 16 and the 126 the SSB 
of $SO(10)$ to $G_{SM}$
occurs in two steps, but
the unification of the $G_{SM}$
couplings occurs in one step,
since the intermediate symmetry group in such cases
is the simple group $SU(5)$.)
Concerning the 45 representation, one can 
show that the only non-trivial positive
definite invariant with degree $\leq 4$ 
(as necessary in order to have a renormalizable potential) 
that can be constructed with the 45
is not extremized  
along any of the two
$G_{SM}$ singlets of the 45.
Moreover, the $G_{SM}$ singlet
of the 210 which has
little group 
$SU(3)_c \otimes SU(2)_L \otimes U(1)_{T_{3R}} \otimes U(1)_{B-L}$
is not a good candidate for the first SSB step
because of the above mentioned need for an intermediate
symmetry group containing
either {\small $SU(4)_{PS}$} or {\small $SU(2)_R$}.

The previous considerations lead to four 
scenarios, in which the first steps of breaking are:

{\footnotesize
\[
  \begin{array}{ll}
    ~(Ia)~~~ SO(10) ~ \buildrel 210 \over \longrightarrow   & \!\!
SU(3)_c\otimes SU(2)_L\otimes SU(2)_R\otimes U(1)_{B-L}\times D   \\ \\
    ~(Ib)~~~ SO(10) ~ \buildrel 210 \over \longrightarrow   & \!\!
SU(4)_{PS} \otimes SU(2)_L \otimes SU(2)_R          \\ \\
    ~(Ic)~~~ SO(10) ~ \buildrel 210 \over \longrightarrow   & \!\!
SU(3)_c\otimes SU(2)_L\otimes SU(2)_R\otimes U(1)_{B-L} \\ \\
    ~(II)~~~ SO(10) ~ \buildrel 54 \over \longrightarrow   & \!\!
SU(4)_{PS} \otimes SU(2)_L \otimes SU(2)_R\times D ~, 
  \end{array}
\]
}

The type-I $SO(10)$ models require that 
an appropriate vector\footnote{Notably,
whereas the vectors used for the
type-Ia and type-Ib models
correspond to maximal little groups,
the vector used for
the type-Ic model, which can be freely chosen within a two-dimensional
space, corresponds to a little group which is not maximal.
The fact that there exist\cite{abrs,gactesi,ofeliatesi,lone}
Higgs potentials constructed with the 210
that have absolute minimum appropriate for the type-Ic case,
{\it i.e.} along a direction which does not correspond to
a maximal little group,
provides one of the few known counter examples of Michelle's conjecture.}
in the space of $G_{SM}$ singlets of the 210
acquires a
v.e.v. (vacuum expectation value)
at the GUT scale.
Analogously the type-II model 
requires that 
the $G_{SM}$ singlet of the 54 acquires a v.e.v.
at the GUT scale.

An appealing\cite{lone} possibility for the completion of
the models of type-Ia,b,c and type-II
is the one of
realizing the second SSB step, at a scale $M_R$,
with the $G_{SM}$ singlet of 
the $126\oplus\overline{126}$ representation, and the third
SSB step
with a combination 
of the {\small $SU(3)_c\otimes U(1)_{e.m.}$}-invariant
vectors of two 10's, in such a way to 
avoid the unwanted relation $m_t=m_b$ \cite{lone}.
Through the see-saw mechanism, the scale $M_R$ is 
related to the masses of
the (almost) left-handed neutrinos. 

The type-Ia,b,c and type-II
models have been studied in the 
Refs.\cite{abrs,gactesi,ofeliatesi,lone,lore,lui,tiz,venezia}, 
and they
have been found to be consistent with the unification of couplings
and
the experimental bound on the
proton lifetime,
although in the case of the type-II model
the consistence with 
the experimental bound on the
proton lifetime is only marginal\cite{lone}.
Within the see-saw mechanism,
one also finds that these models predict 
masses
for the (almost) left-handed neutrinos in
an interesting range. 

This phenomenology
depends however on the values of the
parameters of the Higgs potential,
which are free inputs of the model.
Most importantly,
as mentioned above,
the parameters of the Higgs potential
must be chosen so that the desired SSB pattern
is realized.
Although this does not involve a particular
fine tuning\cite{lone,lore,lui,tiz},
it does introduce an element of undesirable arbitrariness
in the models.
Similarly, the ``matter ingredients" of the models
({\it e.g.}, in the type-I models,
$16 \oplus 16 \oplus 16$ for the fermions
plus $10 \oplus 10 \oplus 126 \oplus \overline{126} \oplus 210$
for the Higgs bosons)
is selected with the only constraint of reproducing
observation
({\it i.e.} the matter content is not constrained 
by any requirement of internal consistency of the models).

\section{RG Naturalness of Higgs Parameters}

$~~~~~~$One way to render a GUT more predictive
would be the discovery of a dynamical mechanism
(quasi) fixing the values of the parameters of
the Higgs potential at the GUT scale $M_X$.
In this section I discuss one such mechanism which
might be available when looking at the GUT as an
effective low-energy description
of a more fundamental theory 
(possibly including gravity).
From this effective theory viewpoint
the GUT becomes relevant for the description
of particle physics at some scale $M^*$ higher 
than $M_X$, and one could investigate
the RGE running of the parameters of the Higgs potential
between $M^*$ and $M_X$.
In certain circumstances the running can {\it push}
the parameters toward certain ranges of values,
and this,
besides having implications for the mass spectrum,
might determine the SSB.
Obviously, in such a scenario the IR structure
of the RGEs would be important.
For example,
to clarify the concept,
let me assume that the RGEs have an IR fixed point
and the running between $M^*$ and $M_X$
is extremely fast.
Under these ideal conditions, the value of the Higgs parameters
at the GUT scale would ``necessarily''
(unless the input parameters at the scale $M^*$
are fine-tuned to avoid that)
be within a small
neighborhood of the IR fixed point.
The SSB would be then essentially determined to be the one
corresponding to the 
values taken by the Higgs parameters at the IR fixed point,
so in this sense it would be dynamically determined.
Clearly one should not expect to find
this ideal scenario in physically relevant contexts.
In realistic cases there might or might not be an IR fixed point
and the running between $M^*$ and $M_X$
might or might not
be fast enough to {\it drive} the parameters
close to their IR destination.
Concerning the speed of the running it must be noted
that this speed can receive important contributions
from the many particles that become effectively massless
at $M_X$, and indeed recent studies\cite{grahamfast}
have found examples of very fast running above $M_X$.
This RG running will not always be driven by
IR fixed points, but one can expect the running
to be characterized by some IR features
and this can affect the likelyhood
for the Higgs parameters to have given values at the GUT scale $M_X$
in corrispondence to given input values at the scale $M^*$.
Importantly, the IR features of the RGEs
will likely be related to symmetries,
so that they might favor values of the Higgs parameters
that correspond to one type of SSB (residual symmetry)
of the model. (This is clarified in the discussion of some
examples in Subsections 3.2 and 3.3.)
One can therefore use the investigation
of a candidate GUT from this RG viewpoint to establish whether
the phenomenological SSB pattern
(the SSB pattern ultimately consistent with $SM$)
is {\it natural} in that GUT.

I observe that, besides leading to the possibility of an increased
predictivity (in the sense clarified above)
of the SSB pattern, viewing GUTs as effective
low-energy
descriptions of a more fundamental theory,
with the associated RG implications,
requires a modification of 
the conventional tests of the naturalness of a GUT.
These conventional tests typically assign a ``low grade"
to GUTs in which a fine tuning of the Higgs parameters is needed
for a phenomenological SSB pattern;
however, the effective-theory viewpoint on GUTs
would require to check whether the
phenomenological SSB pattern
corresponds to fine tuning
of the Higgs parameters at the scale $M^*$.
It is plausible that a scenario requiring no fine tuning
of the Higgs parameters at $M^*$
might correspond via the RG running (for example in presence of
an appropriate infra-red fixed point) to a narrow
region (apparent fine tuning) of the Higgs parameter space
at $M_X$, where the SSB is decided.
On the other hand, it is also plausible
that a SSB pattern corresponding to a significant
portion of the Higgs parameter space
actually requires some level of fine tuning at $M^*$
(for example, the considered portion of Higgs parameter space
might be ``disfavored" by the RG running).

Concerning the scale $M^*$
at which the GUT becomes relevant as an effective
low-energy theory, it should be noticed that,
while any scenario with $M^* \! > \! M_X$
is plausible,
the present (however limited)
understanding of physics beyond the GUT scale $M_X$
suggests that $M^*$ could be within a few orders of magnitude of
the Planck scale $M_P$.
In fact, it is reasonable to expect that beyond the GUT there is
a theory incorporating gravity (a quantum gravity),
and $M_P$ is the scale believed to characterize
this more fundamental theory.

It is also important to realize that the type
of {\it RG naturalness} that I am requiring for GUTs
is really a minimal requirement once the GUT is seen as
an effective low-energy
description of a more fundamental theory.
In order to get a consistent GUT from this viewpoint
one would also want that ``nothing goes wrong"
in going from the scale $M^*$ to the scale $M_X$.
For example, SSB should not occur ``prematurely"
at a scale $\mu_{SSB}$ such that $M_X \! < \! \mu_{SSB} \! < \! M^*$,
and the running of the masses involved in the RGEs
should be taken into account.
In relation to this point, it is interesting
that the investigation
of the {\it RG naturalness} of the parameters of the Higgs
potential might ultimately help understanding also the
emergence of the GUT scale. At present this scale is just
a phenomenological input of a GUT, resulting from the
observed (low-energy) values of the $G_{SM}$ coupling constants,
but it would be interesting to see it emerging as 
a scale within the GUT. By studying the RGEs for
the parameters of the Higgs potential one might find such a
scale; for example, assuming not-too-special initial conditions
for the parameters at the scale $M^*$,
one might find that the running of the parameters is such
that SSB of the GUT occurs typically in the neighborhood
of a certain scale (hopefully a phenomenologically reasonable one).

I also want to stress that one could consider additional
consistency/naturalness conditions in order to render the GUT
consistent with
a working cosmological (early universe evolution) scenario.
Such conditions should be properly formulated in the
language of finite temperature field theory,
and should take into account the fact that (contrary to the expectations
often expressed in the literature) the dependence of couplings on
the renormalization scale is different
from their temperature dependence.

Finally,
before proceeding to the discussion of two GUTs
whose {\it RG naturalness} analysis could be
quite interesting, I would like to point the attention
of the reader toward other ideas expressed in the literature
which are somewhat connected with the {\it RG naturalness}
that I am advocating. 
First of all,
it should be noticed that this {\it RG naturalness}
is closely related to the ideas
on the role 
of the IR structure of the RGEs
in particle physics
discussed in the Refs.\cite{grahamfast,rossold,rossnew}. 
Actually, the observation\cite{rossnew} that
all known physics is consistent
with the idea that the 
relevant (low-energy) parameters 
are strongly influenced by the IR structure of RGEs 
is of encouragement for the hope that the {\it RG naturalness}
that I am advocating might prove useful in analyzing candidate GUTs.
The reader will also notice that there are certain analogies between
the RG techniques useful to investigate {\it RG naturalness}
and those used in studies of stability\cite{orafstab}
and ``finite (SUSY) GUTs''\cite{piguetkazak}.
Concerning stability analyses of the type in Ref.\cite{orafstab},
it is worth
emphasizing that the concept of stability lives
fully at the GUT scale, and therefore does not
involve the interpretation of the GUT as an effective low-energy
description of a more fundamental theory.
Concerning the ``finite GUTs",
I observe that some of them,
those which correspond to an IR fixed point of the RGEs
of the Higgs parameters,
should be expected to appear very {\it RG natural}
from the point of view advocated in this paper;
however, dedicated analyses are necessary
since often the literature
on ``finite GUTs" has not paied much attention to SSB
and certain parameters of the Higgs potential.

\subsection{A nonSUSY Case}

$~~~~~~$In order to render more explicit
the idea of {\it RG naturalness} of the parameters of
the Higgs potential
I want to discuss briefly the Higgs potentials
of two GUTs,
focusing for simplicity on the first step
of SSB at the scale $M_X$. In this subsection
I consider a nonSUSY case, the one of the type-I
$SO(10)$ models reviewed in Sec.2,
the next subsection will be devoted to a SUSY case.

In the type-I
$SO(10)$ GUTs
the first step of SSB
involves the Higgs of the 210-dimensional IRREP.
The most general\cite{abrs,gactesi,ofeliatesi,lone,lore,lui}
(renormalizable) Higgs potential
which can be constructed with the 210
can be written as
\begin{eqnarray}
V(\Phi) &=& A \, ||(\Phi\Phi)_{45}||
+ B \, ||(\Phi\Phi)_{54}||+ C \, ||(\Phi\Phi)_{210}|| 
\nonumber\\
& &+ D \,
\sqrt{||\Phi||} \,
\left((\Phi\Phi)_{210}\times\Phi\right)_1
~,\label{danc}
\end{eqnarray}
where $(R' R'')_R$ stands for the component in the $R$
IRREP of the product of the two IRREPS $R'$ and $R''$.

As discussed in Refs.\cite{abrs,gactesi,ofeliatesi,lone,lore,lui},
many different directions can be
obtained as the minimum
of this Higgs potential 
upon appropriate choices of 
the values of the Higgs parameters $A$, $B$, $C$,
and $D$. Most importantly, different values of the Higgs
parameters, by leading to a different direction for the minimum,
correspond to different groups of intermediate symmetry
(the group of invariance of the vector acquiring a v.e.v.
at the scale $M_X$), and in particular
some of the possible groups of intermediate symmetry
are not phenomenologically interesting since they do not
contain $G_{SM}$.
Following a conventional
approach to GUT model building one would
search for regions of
$A$, $B$, $C$, $D$ parameter space
corresponding to phenomenologically interesting SSB;
for example, some
conditions on the parameters
$A$, $B$, $C$, $D$ 
necessary for the realization of the scenarios of type I-a,b,c
where derived in Refs.\cite{abrs,gactesi,ofeliatesi,lone,lore,lui}.
The analysis of the {\it RG naturalness} (in the sense discussed above)
of these models
would require the study of the RGEs of the
parameters $A$, $B$, $C$, $D$,
and the identification of the regions
of $A$, $B$, $C$, $D$ 
parameter space which are {\it favored} by the running.
For example, the reader can immediately
see the substantial difference between
the (ideal) case in which one finds
a strongly attractive fixed point 
in a region
of $A$, $B$, $C$, $D$ 
parameter space corresponding to the type-Ia scenario
and the (ideal) case in which one finds a similarly
attractive fixed point 
in a region
of $A$, $B$, $C$, $D$ 
parameter space corresponding to a phenomenologically
inconsistent SSB.
In general, even when there are no IR fixed points,
one can expect that regions 
of $A$, $B$, $C$, $D$ 
parameter space corresponding to certain symmetries
will be preferred by the infrared structure of the RGEs
with respect to
regions 
of $A$, $B$, $C$, $D$ 
parameter space corresponding to other symmetries.

\noindent
The investigation of the relevant RGEs
is left for future work.

\subsection{A SUSY Case}

$~~~~~~$In order to illustrate
the idea of {\it RG naturalness} 
also in the context of SUSY GUTs,
in this subsection
I want to discuss briefly the Higgs potential
of the simplest such model: minimal SUSY$SU(5)$.
While nonSUSY $SU(5)$ GUTs are essentially ruled out
experimentally (in particular, they predict one-step unification),
and therefore the $SO(10)$ models discussed in the previous subsection
and in Sec.2 provide a somewhat minimal (simple-group based)
phenomenologically interesting nonSUSY GUT scenario,
on the SUSY side 
even the $SU(5)$ case (while being affected by several technical
difficulties\cite{flip2}, 
which are better handled by some of its extensions\cite{flip1})
can be made consistent
with the low-energy data on the coupling constants and
the present experimental lower limit 
on proton decay.
It is therefore
reasonable to use the minimal SUSY$SU(5)$ model,
in which the chiral field $\Sigma$ involved in 
the first step of SSB
is described by the 24-dimensional (adjoint) IRREP,
to illustrate
the idea of {\it RG naturalness} 
in the context of SUSY GUTs.

The 
Higgs potential relevant 
for the first step of SSB
can be written (see, {\it e.g.},
Ref.\cite{susysu5})
as
\begin{eqnarray}
V = V_{susy} 
+ V_{soft}
~,\label{pot}
\end{eqnarray}
where $V_{susy}$ is the supersymmetric part of the potential,
which can be expressed in terms of the superpotential $W$,
\begin{eqnarray}
W(\Sigma) = {\mu \over 2} Tr \Sigma^2 + {b \over 3} Tr \Sigma^3
~,\label{superpot}
\end{eqnarray}
as
\begin{eqnarray}
V_{susy} (\Sigma) = \huge{|} {\partial W \over \partial \Sigma_i} \huge{|}^2
+ \rm{``D terms''}
~,\label{susypot}
\end{eqnarray}
while $V_{soft}$ contains the terms breaking SUSY ``softly''
and can be written as\cite{susysu5}
\begin{eqnarray}
V_{soft}(\Sigma) = {m_o^2 \over 2} \, |\Sigma_i|^2
+ m_o \, \Sigma_i \, {\partial W \over \partial \Sigma_i} 
+ m_o \, (A-3) \, W + \rm{h.c.}
~.\label{softpot}
\end{eqnarray}

The supersymmetric part of the potential
has several degenerate minima\cite{susysu5,buccellasusy},
corresponding to directions
with little group
$SU(5)$, $SU(4) \otimes U(1)$, or $G_{SM}$.
The degeneracy is removed by the
soft terms; in particular, in the phenomenologically relevant 
case $m_o / \mu \! < \! < \! 1$
one can show\cite{susysu5} that
for $A \! < \! 3$ the minimum is in a 
direction
with little group $G_{SM}$ (SSB),
whereas for $A \! > \! 3$ the minimum is in a 
direction
with little group $SU(5)$ (unbroken symmetry),

Within this SUSY model
the analysis 
of the {\it RG naturalness} of the phenomenological SSB
to $G_{SM}$ 
requires the study of the RGEs 
of the
parameters of $V$
in order to establish whether 
it is likely/plausible that 
$A \! < \! 3$ at $M_X$.
Ideally, one would like to find a
strongly attractive infra-red fixed point
corresponding to $A \! < \! 3$.

\section{A Simple GUT Matter Content ?}

$~~~~~~$As illustrated by the review of SO(10) GUTs in Sec.2,
GUTs typically involve a remarkably complex matter content.
Most notably, the Higgs sector consists of several carefully
selected\footnote{The reader should notice that I put the emphasis
on the (apparently) {\it ad hoc}
selection of the Higgs representations,
rather than the size of the representations involved.
For example, the Higgs content of the SO(10) GUTs reviewed
in Sec.2 is unappealing to
some authors primarily because of the size
of some of the IRREPs involved, while I am most puzzled by the
fact that among the available IRREPs only a carefully
selected bunch
is to be involved
(in order to achieve agreement with phenomenology).}
IRREPs of the GUT group,
and, like in the $SM$,
the fermionic sector of leptons and quarks is
arbitrary\footnote{For GUTs in which (like the $SM$
and unlike SO(10) GUTs) the absence of anomaly
results from an ``accidental''
cancellation among the anomaly contributions of the
different IRREPs involved,
one has even more reasons to be puzzled.} 
and is plagued by the family triplication.

This complexity might well be telling us that the GUT idea
needs drastic revisions; however,
in this paper I take the point of view
that the complexity of the matter content might be only apparent.
I therefore want to mention a few appealing scenarios in which
this complexity might arise from a fundamental simplicity.

For continuity with the line of argument advocated in the previous
section, let me start by mentioning the possibility that
as an effective low-energy description of a more fundamental theory,
the (effective)
matter content of the GUT at the scale $M_X$
might be fixed by the RG running.
It is in fact plausible that some IRREPs tend to get 
heavy masses via RG running, whereas the masses of other 
IRREPs (the ones relevant for symmetry breaking
and low-energy phenomenology)
might tend to be light ({\it i.e.} order $M_X$ or less).
This type of running of masses (or other parameters) might 
even be responsible for the cancellation of anomalies
at low energies; indeed, the RG running is known to
be easily driven by symmetries.

\noindent
The viability of such a
mechanism of selection of the matter content could be tested
in simplified/familiar contexts; for example, one could study
the RG equations between $M_P$ and $M_X$ in a model obtained
by adding one extra IRREP 
of Higgses to the type-I SO(10) GUT, 
and check whether the extra IRREP tends to decouple from the rest.

If not resulting from RG running,
the complexity of the matter content might even
be the result of compositeness, in line
with the most honored tradition of particle physics.
I point to the attention of the reader
the investigations
reported in Ref.\cite{preons}
(and references therein)
which consider the possibility of
preonic GUTs.

Another hypothesis which has been gaining some momentum in the
literature on low-dimensional {\it Quantum-Gravity} 
toy models 
is that {\it Quantum Gravity} 
might be quite selective concerning the
type of matter ``it likes to deal with'', {\it i.e.} the requirement
of overall consistency of {\it Quantum Gravity} might fix the matter
content.
Results pointing (however faintly)
in this direction can be found for example in 
certain studies
of discretized 
two-dimensional
{\it Quantum Gravities}\cite{triangulazio},
and studies of the Dirac quantization of certain two-dimensional
{\it Quantum Gravities}
in the continuum\cite{gacjackpapers}.

In general it is clear that, 
in trying to make sense
of the matter content of a GUT,
it would be useful 
to observe group theoretical
properties that single out the IRREPs
corresponding to that matter content.
Just to give an example of what I mean
with ``group theoretical properties'',
I observe that the following SO(10) relations

\begin{eqnarray}
~ 16 \otimes 16 &=& 10_S \oplus 120_A \oplus 126_S
\nonumber\\
~ \overline{16} \otimes \overline{16} &=&
10_S  \oplus 120_A  \oplus \overline{126}_S
\nonumber\\
~ \overline{16} \otimes 16 &=&
1 \oplus 45 \oplus 210
\nonumber\\
~ \overline{16} \otimes 16 \otimes 16 &=&
16 \oplus 16 \oplus 16 \oplus 144 \oplus 144
\nonumber\\
&&\oplus 560
\oplus 560
\oplus 1200
\oplus 1440
\label{bareobs}
\end{eqnarray}
come quite close to singling out the matter
content of the type-I SO(10) GUTs.
One could then examine similar group theoretical properties,
and try to figure out their origin.
For example,
an observation such as (\ref{bareobs})
might motivate related work on preonic GUTs,
and from the preonic viewpoint one might
be even more intrigued by the observation
that the fermions (16's)
would be contained in
the product of an odd number of preons
($\overline{16} \otimes  16  \otimes  16$),
while the bosons (Higgses 
of $10 \oplus 126 \oplus \overline{126}  \oplus 210$)
would be contained in 
the product of an even number of preons
($(16 \otimes 16) \oplus (\overline{16} \otimes  \overline{16})
\oplus (\overline{16}  \otimes  16)$).

\section{Closing Remarks}

$~~~~~~$Perhaps the only 
robust concept discussed in this paper is the one concerning
the way in which the conventional tests of the naturalness
of a GUT need to be modified if the GUT is seen
as a low-energy effective description of a more fundamental theory.

On a more speculative side,
I also articulated the hope that the correct GUT (if there is one)
could be such that its SSB pattern is essentially
predicted (in the sense of the {\it RG naturalness}
I discussed) rather than being a free input;
this would fit well the general trend of increased
predictivity at each new stage of our understanding
of particle physics.

I have also looked at the complexity of the matter content
of GUTs, and explored the possibility that this might be
an apparent complexity, hiding a fundamental simplicity.
In Sec.4 I have speculated on a few appealing candidates
for this simplicity; however, it is reasonable to expect that
real progress in this direction
will require dramatic new
developments.

\bigskip
\bigskip
\bigskip
The ideas discussed in this paper have been developing over
a long period of time, and there are
several colleagues I am indebted to.
Subsec.3.3 was added only after the start of preliminary
discussions with B. Allanach and
O. Philipsen, ultimately leading to the
collaboration\cite{nextgut}. The suitability of 
the minimal SUSYSU(5) model 
for a {\it RG naturalness} analysis emerged in the context of those
discussions.
I would also like to thank G. Ross, for exposing me
to his powerful arguments
on the possible role of the infrared structure
of renormalization group equations
in the understanding of particle physics,
and J.D. Bjorken, for stimulating feed-back on Sec.4.
Finally, I am greatly indebted to
F. Buccella, who, valiantly assisted by L. Rosa,
guided from his vantage point my first GUTsy steps\cite{gactesi}
a few years ago.


\begin{thebibliography}{9}

\bibitem{amal} U. Amaldi, W. de Boer, and H. Furstenau, Phys. Lett. B260
(1991) 447.

\bibitem{padb} Review of Particle Properties, Phys. Rev. D45 (1992).

\bibitem{oliv} A. le Yaouanc, L. Oliver, O. P\`ene, and J. C. Raynal, 
Phys. Lett. B72 (1977) 53.

\bibitem{frmi} H. Georgi, Particles and Fields, C. E. Carlson, AJP, 1975; 
H. Fritzsch and P. Minkowski, Ann. of Phys. 93 (1975) 183.

\bibitem{kuzm} V. Kuzmin and N. Shaposhnikov, Phys. Lett. B92 (1980) 115.

\bibitem{ext45} L. Li, Phys. Rev. D9 (1974) 1723.

\bibitem{abrs} M. Abud, F. Buccella, L. Rosa, A. Sciarrino, Zeitschrift 
$f\ddot u$r Phys. C44 (1989) 589; F. Buccella and L. Rosa, Nucl. Phys. B 
(Proc. Suppl.) 28A (1992) 168.

\bibitem{gactesi} G. Amelino-Camelia, ``Spontaneous Symmetry Breaking in 
SO(10) GUT Models with 
\indent $\tau_p\sim 10^{33}$ years", {\it Laurea Thesis}
(1990).

\bibitem{ofeliatesi} O. Pisanti, ``An Analysis of SO(10) Models with 
Intermediate Symmetry $SU(3)_c\otimes 
\indent
SU(2)_L \otimes SU(2)_R \otimes 
U(1)_{B-L}$", {\it Laurea Thesis} (1992).

\bibitem{lone} F. Acampora, G. Amelino-Camelia, F. Buccella, O. Pisanti, 
L. Rosa, and T. Tuzi, 
Il Nuovo Cimento A108 (1995) 375.

\bibitem{lore} Q. Shafi and C. Wetterich, Phys. Lett. B 85 (1979) 52; 
F. Buccella, L. Cocco, and C. Wetterich, Nucl. Phys. B 243 (1984) 273; 
M. Abud, F. Buccella , A. Della Selva, A. Sciarrino, R. Fiore, and G. 
Immirzi, Nucl. Phys. B263 (1986) 336.

\bibitem{lui} F. Buccella and L. Rosa, Zeitschrift f$\ddot u$r Phys. C36 
(1987) 425.

\bibitem{tiz} D. Chang, R. N. Mohapatra, and M. K. Parida, Phys. Rev. Lett.
52 (1984) 1072; D. Chang, J. M. Gipson, R. E. Marshak, R. N. Mohapatra, and 
M. K. Parida, Phys. Rev. D31 (1985) 1718; F. Buccella, L. Cocco, A. Sciarrino,
and T. Tuzi, Nucl. Phys. B274 (1986) 559; R. N. Mohapatra and M. K. Parida, 
Phys. Rev. D47 (1993) 264.

\bibitem{venezia} G, Amelino-Camelia, F. Buccella, and L. Rosa,
hep-ph/9405334,
in the proceedings of {\it 2nd International 
Workshop on Neutrino Telescopes}, Venezia, Italy, 1990.

\bibitem{grahamfast} M. Lanzagorta and G.G. Ross,
Phys.Lett. B349 (1995) 319.

\bibitem{rossold} B. Pendleton and G.G. Ross,
Phys. Lett. B98 (1981) 291.

\bibitem{rossnew} M. Lanzagorta and G.G. Ross,
Phys. Lett. B364 (1995) 163;
G.G. Ross, Phys. Lett. B364 (1995) 216. 

\bibitem{orafstab} See, for example,
the paper by
Abud, Buccella , Della Selva, Sciarrino, Fiore, and  
Immirzi, in \cite{lore},
and J. Basecq, S. Meljanac,
and L. O'Raifeartaigh, Phys. Rev. D39 (1989) 3110. 

\bibitem{piguetkazak} See, for example,
O. Piguet, hep-th/9606045,
to appear in the proceedings of {\it 10th International
Conference on Problems of Quantum Field Theory}, Alushta, 
Ukraine, 1996,
and
D.I. Kazakov, M.Yu. Kalmykov, I.N. Kondrashuk, 
and A.V. Gladyshev, Nucl. Phys. B471 (1996) 389.

\bibitem{flip2} J.L. Lopez and D.V. Nanopoulos,
hep-ph/9511266,
presented to {\it Int. School on Subnuclear Physics, 
Vacuum and Vacua}, Erice, Italy, 1995. 

\bibitem{flip1} I. Antoniadis, J. Ellis, J.S. Hagelin, 
and D.V. Nanopoulos,
Phys. Lett. B194 (1987) 231.

\bibitem{susysu5} G.A. Diamandis and B.C. Georgalas, 
Phys. Lett. B160 ( 1985) 263;
M. Drees, S. Meljanac,
M. Milekovic, and S. Pallua,
Phys. Lett. B178 (1986) 226.

\bibitem{buccellasusy} F. Buccella, J.P. Derendinger, S. Ferrara, C.A. Savoy,
Phys. Lett. B115 (1982) 375.

\bibitem{preons} J.C. Pati, in the proceedings of {\it High 
Energy Physics and Cosmology}, Trieste, Italy, 1993.

\bibitem{triangulazio} J. Ambjorn, in the proceedings of {\it Fluctuating 
Geometries in Statistical Mechanics and Field Theory},
Les Houches (Session LXII), France, 1994.

\bibitem{gacjackpapers} D. Cangemi and R. Jackiw, 
Phys. Lett. B337 (1994) 271;
G. Amelino-Camelia, D. Bak, 
and D. Seminara, MIT-CTP-2539/OUTP-96-18P/SNUTP-96-044
Phys. Rev. D54 (1996) in press.

\bibitem{nextgut} B. Allanach, G. Amelino-Camelia, 
and O. Philipsen, in preparation.

\end{thebibliography}
\end{document}